\documentclass[%superscriptaddress,
reprint,
nofootinbib,
 amsmath,amssymb,
 aps,
]{revtex4-2}
\usepackage{orcidlink}

\usepackage[utf8]{inputenc}
\usepackage{xcolor}
\usepackage{booktabs}

\usepackage{graphicx}% Include figure files
\usepackage{dcolumn}% Align table columns on decimal point
\usepackage{bm}% bold math
\usepackage{amsmath}
\usepackage{braket}
\usepackage{natbib}
\usepackage[version=4]{mhchem}

\usepackage{listings}

\makeatletter

\usepackage{ulem}
\usepackage{siunitx}
\usepackage{graphicx,multirow}
\usepackage{array}
\usepackage{glossaries}
\usepackage{tikz}
\usepackage[caption=false]{subfig}

\newacronym{vqe}{VQE}{variational quantum eigensolver}
\newacronym{sapt}{SAPT}{symmetry-adapted perturbation theory}
\newacronym{pdb}{PDB}{protein data bank}
\newacronym{dft}{DFT}{density functional theory}
\newacronym{hf}{HF}{Hartree-Fock}
\newacronym{mo}{MO}{molecular orbital}
\newacronym{ao}{AO}{atomic orbital}
\newacronym{shci}{SHCI}{semistochastic heat bath configuration interaction}
\newacronym{nor}{NOR}{nitric oxide reductase}
\newacronym{fci}{FCI}{full configuration interaction}
\newacronym{1pdm}{1-PDM}{one-particle density matrix}
\newacronym{2pdm}{2-PDM}{two-particle density matrix}
\newacronym{gpu}{GPU}{graphics processing unit}
\newacronym{casscf}{CASSCF}{complete active space self-consistent field}

\newcommand{\quotes}[1]{``#1''}
\newcommand{\Eelst}{E_{\text{elst}}}
\newcommand{\JJ}{\mathcal{J}}
\newcommand{\R}{\mathbb{R}}
\newcommand{\td}{\text{d}}

\begin{document}

\title{Estimation of electrostatic interaction energies on a trapped-ion quantum computer}

\date{2023}
%-------------------------------------------------------
% author list
\author{Pauline J. Ollitrault~\orcidlink{0000-0003-1351-7546}}\email{pauline.ollitrault@qcware.com}

\author{Matthias Loipersberger~\orcidlink{0000-0002-3648-0101}}

\author{Robert M. Parrish~\orcidlink{0000-0002-2406-4741}}
\affiliation{QC Ware Corp, Palo Alto, USA and Paris, France}

\author{Alexander Erhard~\orcidlink{0000-0002-5020-2271}}
\author{Christine Maier~\orcidlink{0000-0001-5633-8607}}
\author{Christian Sommer~\orcidlink{0000-0002-8900-2744}}
\author{Juris Ulmanis~\orcidlink{0000-0003-0015-5371}}
\author{Thomas Monz~\orcidlink{0000-0001-7410-4804}}\altaffiliation[Also at ]{Universit{\"a}t Innsbruck, Institut f{\"u}r Experimentalphysik, Innsbruck, Austria}
\affiliation{Alpine Quantum Technologies GmbH, 6020 Innsbruck, Austria}

\author{Christian Gogolin~\orcidlink{0000-0003-0290-4698}}
\affiliation{Covestro Deutschland AG, 51373 Leverkusen, Germany}

\author{Christofer S.\ Tautermann~\orcidlink{0000-0002-6935-6940}}\altaffiliation[Also at ]{University of Innsbruck, Department of General, Inorganic and Theoretical Chemistry, Innsbruck, Austria}
\affiliation{Medicinal Chemistry, Boehringer Ingelheim Pharma GmbH \& Co. KG, 88397 Biberach, Germany}

\author{Gian-Luca R. Anselmetti\orcidlink{0000-0002-8073-3567}}
\author{Matthias Degroote~\orcidlink{0000-0002-8850-7708}}
\author{Nikolaj Moll~\orcidlink{0000-0001-5645-4667}}
\author{Raffaele Santagati~\orcidlink{0000-0001-9645-0580}}
\author{Michael Streif~\orcidlink{0000-0002-7509-4748}}
\email{michael.streif@boehringer-ingelheim.com}
\affiliation{Quantum Lab, Boehringer Ingelheim, 55218 Ingelheim am Rhein, Germany}

%-------------------------------------------------------
\begin{abstract}
We present the first hardware implementation of electrostatic interaction energies using a trapped-ion quantum computer. As test system for our computation, we focus on the reduction of \ce{NO} to \ce{N2O} catalyzed by a nitric oxide reductase (NOR). The quantum computer is used to generate an approximate ground state within the NOR active space. To efficiently measure the necessary one-particle density matrices, we incorporate fermionic basis rotations into the quantum circuit without extending the circuit length, laying the groundwork for further efficient measurement routines using factorizations. Measurements in the computational basis are then used as inputs for computing the electrostatic interaction energies on a classical computer.  
Our experimental results strongly agree with classical noise-less simulations of the same circuits, finding electrostatic interaction energies within chemical accuracy despite hardware noise.
This work shows that algorithms tailored to specific observables of interest, such as interaction energies, may require significantly fewer quantum resources than individual ground state energies would in the straightforward supermolecular approach. 

\end{abstract}
%-------------------------------------------------------
\maketitle

\section{Introduction}

The potential of quantum computers to simulate molecules is promising for many applications in industry and research~\cite{santagati2023drug, dalzell2023quantum,cao2018potential}. 
While quantum computers which could simulate classically intractable molecules yet have to be built, the task of identifying relevant industrial applications of these future machines is equally crucial. This goal involves not only identifying classically challenging molecules that could benefit from quantum computations~\cite{reiher2017elucidating,goings2022reliably,rubin2023fault, kim2022fault}, but also developing quantum algorithms which perform the computations required for specific applications in industry. 

To date, most quantum algorithms have been developed to capture more accurate ground state total energies~\cite{peruzzo2014variational, von2021quantum}. However, accurate computations of properties beyond ground state energy are essential for many industrial applications~\cite{knill2007optimal, steudtner2023fault, o2022efficient}. A prime example is the estimation of interaction energies between two molecules, a critical first step in computational drug design for ranking the efficacy of ligands against a target such as a protein~\cite{palermo2014computational}. 
On a quantum computer, interaction energies could be obtained through three distinct ground state energy computations: one for each molecule and one for the entire system. Given that interaction energies typically constitute only a fraction of the total ground state energies, this method —known as the supermolecular approach— requires a high precision in  each ground state energy calculation~\cite{gutowski1986basis,gutowski1987proper}.

\Gls{sapt} offers an alternative approach by expressing the interaction energy from a perturbative treatment of the intermolecular potential~\cite{jeziorski1976first, jeziorski1994perturbation, patkowski2020recent}. Recent theoretical advancements have laid the foundation for harnessing emerging quantum computing capabilities to compute more accurate interaction energies via \gls{sapt}~\cite{malone2022towards,loipersberger2023accurate}. Specifically, it was demonstrated that a near-term quantum algorithm, the \gls{vqe}~\cite{peruzzo2014variational}, could enhance the accuracy of SAPT calculations and provide an alternative to the supermolecular approach. However, this was only tested via the emulation of the quantum computer on classical computers.

In this work, we present the first experimental demonstration toward the computation of \gls{sapt} energies on a quantum computer, focusing specifically on the quantum computation of the electrostatic energy. We note that the computation of the electrostatic term by itself does not require symmetry adaption. However, we still refer to \gls{sapt} interaction energies as this work presents the first step toward the goal of implementing the full second order \gls{sapt} energies on quantum hardware~\cite{malone2022towards, loipersberger2023accurate}.

An important part of this quantum algorithm is its embedding into a classical framework. Therefore, it is desirable to apply it to molecules whose sizes and complexities are representatives of standard problems in industrial use cases. 
This allows to resolve any classical bottleneck, ensuring a smooth transition to an era with improved quantum hardware.
Hence, as a test system for our experiments, we focus on the active site of a \gls{nor}, a key player in the chemical reaction that reduces nitric oxide (\ce{NO}) to nitrous oxide (\ce{N2O}), i.e. an important step of the nitrogen cycle. Several intermediates in the catalytic cycle have electronic structures that are strongly correlated and thus challenging to simulate with classical methods. As a result, the \gls{nor} active site serves as a valuable benchmark system for quantum algorithms. Notably, \gls{nor} is a member of the cytochrome P450 superfamily, renowned for its monoxygenases that play a central role in drug metabolism. The fact that \gls{nor} belongs to the P450 superfamily renders the exploration of the catalytic cycle of NO reduction not only intrinsically valuable but also broadly significant due to its structural parallels with other P450 enzymes. 

In order to make our computations possible on today's quantum computers with limited qubit counts and gate fidelities, we rely on classical quantum chemistry methods to preprocess and heavily simplify the system. Specifically, while we treated the studied molecules with more than a thousand orbitals classically, we only utilize four orbitals (mapped to eight qubits) to account for the molecule's strong correlation on the quantum computer. Despite the quantum algorithm only modeling a very small part of the system, a task that could be readily handled by any classical computer, our work serves as the first demonstration of such a computation on quantum hardware. 
An interesting feature emerging from this experiment is the measurement in a single basis allowed by the implementation of a fermionic basis change in the quantum circuit. The same circuit primitives could also be used to measure \gls{2pdm} with the help of double factorization~\cite{oumarou2023accelerating, hohenstein2023efficient, von2021quantum}, greatly reducing the number of measurement circuits compared to the \quotes{na\"ive} approach. This feature highlights our work also as a demonstration of experimental capabilities relevant beyond the \gls{sapt} framework.

In Sec.~\ref{sec:sapt_elec} we review the \gls{sapt} method and explain how to efficiently compute the first term of its perturbation series, the electrostatic contribution. Next, in Sec.~\ref{sec:classical}, we highlight the crucial classical steps that account for the preparation of the molecular systems and the optimization of the quantum circuits. Finally, in Sec.~\ref{sec:results}, we review the implementation of the quantum circuits on quantum hardware and discuss the experimental results.

\section{Quantum computations of the electrostatic energy}
\label{sec:sapt_elec}

The interaction energy between two monomers $A$ and $B$ is given in the supermolecular approach by
\begin{align}
E_\mathrm{int} = E_{AB} - E_A - E_B\,,
\label{eq:interaction_energies}
\end{align}
where $E_{AB}$, $E_{A}$, and $E_{B}$ denote the ground state energies of the dimer, monomer $A$, and monomer $B$, respectively. Consequently, using this approach, the computation of the interaction energy requires three separate ground-state energy calculations, one of which is performed on the larger dimer system.
An alternative approach to computing interaction energies is \gls{sapt} \cite{jeziorski1976first, jeziorski1994perturbation}. In \gls{sapt}, the interaction between the two molecules is treated as a perturbation, and the solution of the symmetrized Rayleigh-Schrödinger equation yields the interaction energy as a perturbation series, written in \textit{polarization} terms, $E_\mathrm{pol}$ and \textit{exchange} terms, $E_\mathrm{exch}$, as
\begin{equation}
    E_\mathrm{int} = \sum_{n}(E_\mathrm{pol}^{(n)} + E_\mathrm{exch}^{(n)}).
    \label{eq:sapt}
\end{equation}
where $n$ is the perturbation order.

The polarization and exchange terms of Eq.~(\ref{eq:sapt}) are commonly relabelled to their interaction energy kind. 
The four main interaction types are the electrostatic energy,  the exchange energy,  the induction energy and the dispersion energy.
The exact expression of these main terms depends on the level of truncation in the chosen \gls{sapt} method.
We direct the reader to Refs.~\cite{parker2014levels,schriber2023levels} for a more detailed introduction to \gls{sapt}.
It is important to remark here that on top of providing insight to the underlying forces of the intermolecular interaction, this approach also eliminates the need for a ground state energy calculation of the dimer system and the propagation of large errors from energy calculations into typically small energy differences.

In this study, we focus on the quantum computation of a single interaction type, $E_\mathrm{elst}$, which describes the electrostatic energy between the two monomers. 
We employ the \gls{fci} \gls{sapt} formalism of Ref.~\cite{korona2008first} wherein a complete (or \gls{vqe}-type near-complete) treatment of electron correlation is assumed at each level of intermolecular perturbation.
In this case, the electrostatic energy can be calculated via
\begin{equation}
    \Eelst = \sum_{pp'} \JJ_{pp'}^B \gamma^A_{pp'}
    \label{eq:Eelst_main}
\end{equation}
where $\gamma^A_{pp'}$ is the spin-summed \gls{1pdm} of monomer $A$ in a (non-orthogonal) spin-restricted spatial \quotes{\gls{ao}} basis $\{\phi_p(\vec{r})\}$
and
\begin{equation}
\begin{split}
    \JJ_{pp'}^B =& \sum_{qq'} (pp'|qq') \gamma^B_{qq'} + \frac{1}{N_A} \sum_{qq'} \gamma^B_{qq'}V^A_{qq'} S_{pp'} \\ + &\frac{1}{N_B} \sum_{qq'} \gamma^B_{qq'}S_{qq'} V^B_{pp'}  
    +\frac{V_{AB}}{N_A N_B} \sum_{qq'} \gamma^B_{qq'} S_{qq'} S_{pp'}
    \end{split}
    \label{eq:Jb_tensor}
\end{equation}
is the generalized electrostatic potential matrix.
Here, $N_A$ and $N_B$ are the numbers of electrons of monomer $A$ and monomer $B$ respectively. The first term comprises the two-body integrals, $(pp'|qq')$ and the spin-summed \gls{1pdm} of monomer $B$, $\gamma^B_{qq'}$.
In the second and third terms, $V^{X}_{pp'}$ is the nuclear potential from monomer $X$ and $S_{pp'}$ the overlap matrix in the \gls{ao} basis. In the last term, $V_{AB}$ is the inter-monomer nuclear repulsion term. All formal definitions can be found in Appendix~\ref{app:Eelst}. In this work, the orbital basis is always defined as the dimer basis.

The expression of Eq.~(\ref{eq:Eelst_main}) is particularly advantageous when monomer $B$ is well described by classical methods such as \gls{hf}, \gls{dft} or other low polynomial-scaling approaches as, in this case, $\JJ^B_{pp'}$ can easily be obtained in classical preprocessing.
The remaining task is to find the (crucial) interaction of the electronic density of $A$ with the full electrostatic potential of $B$. To do so $\gamma_{pp'}^A$ needs to be computed. When the ground state wavefunction of $A$, $\ket{\Psi_A}$, is represented in whole, or in part, on a quantum computer, the use of an orthonormal \quotes{\gls{mo}} spatial basis is commonly required. The MO basis is defined by
\begin{equation}
    \psi_t(\vec{r}_1) \equiv \sum_p C_{pt}\phi_p(\vec{r}_1)
\end{equation}
with the orthonormality condition 
\begin{equation}
    \sum_{pp'} C_{pt}S_{pp'}C_{p't'} = \delta_{tt'} \quad \forall \, t,t'.
\end{equation}
The \glspl{1pdm} in the \gls{ao} and \gls{mo} basis are then related by
\begin{equation}
    \sum_{tt'}C_{pt} \gamma_{tt'}^A C_{pt'} \equiv \gamma_{pp'}^A.
\end{equation}
In the \gls{mo} basis Eq.~(\ref{eq:Eelst_main}) becomes
\begin{equation}
    E_{\text{elst}} = \sum_{tt'} \JJ^B_{tt'} \gamma_{tt'}^A = \sum_{tt'}\sum_{pp'} C_{pt}\JJ^B_{pp'}C_{p't'} \gamma_{tt'}^A.
    \label{eq:Eelst_mo}
\end{equation}
The rewriting of $E_{\text{elst}}$ in the \gls{mo} basis, in Eq.~(\ref{eq:Eelst_mo}), is interesting because it makes apparent that a different set of \glspl{mo}, $\bar{C}_{pv}$ can be chosen such that $\bar{\JJ}^B_{vv'}$ is diagonal, i.e.,
\begin{equation}
    \bar{\JJ}^B_{vv'} \equiv \sum_p \bar{C}_{pv} \JJ^B_{pp'}\bar{C}_{p'v'} = \bar{w}_v\delta_{vv'} \quad \forall \, v, v'.
\end{equation}
This \quotes{electrostatic potential natural orbital basis} is simply obtained by rotating the canonical \glspl{mo} as
\begin{equation}
    \bar{C}_{pv} = \sum_t C_{pt} U_{tv}
\end{equation}
where $U_{tv}$ are the eigenvectors of $\JJ^B_{tt'}$ and $\bar{w}_v$ the corresponding eigenvalues. 
Finally, Eq.~(\ref{eq:Eelst_mo}) becomes
\begin{equation}
    E_{\text{elst}} = \sum_{v} \bar{w}_v\bar{\gamma}^A_{vv} 
    \label{eq:electrostatics_from_rotated_rdm}
\end{equation}
with $\bar{\gamma}^A_{vv} = \sum_{tt'} U_{tv} \gamma_{tt'}^A U_{t'v}$.

Importantly, when the wavefunction is represented on a quantum computer with a spin-restricted Jordan-Wigner formalism, the diagonal elements of the \gls{mo} \gls{1pdm} correspond to commuting, diagonal single-particle Pauli measurements
\begin{equation}
    \bar{\gamma}^A_{vv} = 1 - \bra{\bar{\Psi}^A} \hat{Z}_{v,\alpha}/2\ket{\bar{\Psi}^A} - \bra{\bar{\Psi}^A} \hat{Z}_{v,\beta}/2\ket{\bar{\Psi}^A}
    \label{eq:diag_meas}
\end{equation}
where the Pauli operator $\hat{Z}_{v,\alpha}$ acts on the qubit representing \gls{mo} $v$ with  spin $\alpha$ and $\ket{\bar{\Psi}^A}$ is the ground state of $A$ in the electrostatic potential natural orbital basis. 
Note that the determination of the ground state $\ket{\Psi_A}$ can still be performed as usual with the canonical \glspl{mo} and the transformation to the electrostatic potential natural orbitals can simply be implemented by decomposing the fermionic basis rotation $U_{tv}$ into Givens rotations~\cite{arrazola2022universal,Anselmetti_2021} and appending them to the quantum circuit right before the measurements. 

Ultimately, we see that this procedures allows to obtain $E_{\text{elst}}$ from a single set of measurements since all the operators which expectation values are required to reconstruct $\gamma^A_{vv}$, as defined in Eq.~(\ref{eq:diag_meas}), commute with each other. 
The employed method contrasts with the \quotes{na\"ive} approach where the number of distinct measurement circuits to reconstruct the \gls{1pdm} grows with quadratically with the number of  orbitals $N$. 

In comparison, the supermolecular approach, as per Eq.~(\ref{eq:interaction_energies}), requires the calculation of the ground state energies of the both monomer and dimer systems, which confers a clear advantage to the perturbative alternative. 
Naively, this method requires up to $\mathcal{O}(N^4)$ measurement circuits. 
However, an interesting parallel with the present work is that the same change of fermionic basis approach can be used to measure the \gls{2pdm} with the help of double factorization \cite{oumarou2023accelerating, hohenstein2023efficient, von2021quantum}, reducing the number of measurement circuits. Furthermore randomizing over these basis rotations would allow for a classical shadow approach only scaling with the number of occupied fermionic modes \cite{low22_class_shadow_fermion_with_partic_number_symmet}.

\begin{figure*}[ht]
\centering
\includegraphics[width=0.9\linewidth]{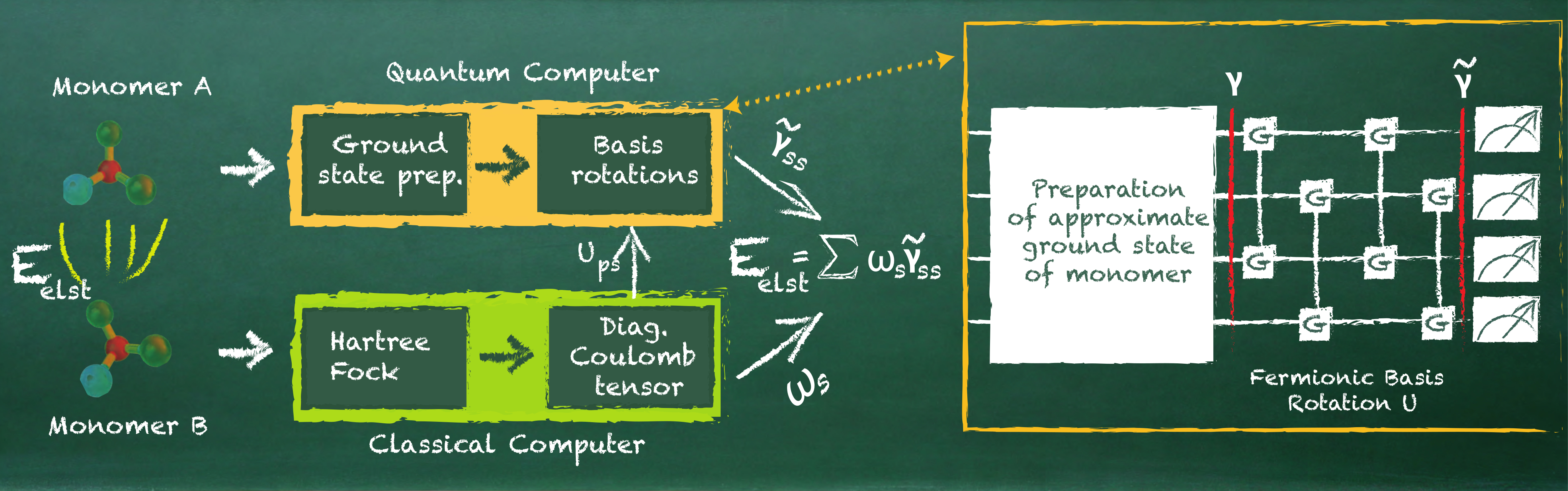}
    \caption{Visualization of the quantum computation of the electrostatic interaction energy. (left) The output of the classical Hartree-Fock calculations is fed into the quantum computer, where a previously generated approximate ground state of monomer A is rotated in the basis given in Eq.~(\ref{eq:electrostatics_from_rotated_rdm}). Measuring the diagonal elements of the 1-RDM allows for an efficient calculation of the electrostatic energy. (right) Visualization of the quantum circuit, including preparing the approximate ground state and the subsequent fermionic basis rotation, see also Fig.~\ref{fig:quantum_circuit}.}
    \label{fig:quantum_sapt}
\end{figure*}

\section{Classical processing}
\label{sec:classical}

As a test system, we exploit the reduction of NO to \ce{N2O} catalysed by \gls{nor}. Details on this reaction, such as the proposed catalytic cycle are discussed in Appendix~\ref{app:molecule}.
Ref.~\cite{shimizu2000proton} discusses the importance of intermolecular interactions in the reaction mechanism. For instance, water molecules surrounding the \gls{nor} active site form a hydrogen bonding network and act as a proton delivery pathway for the formation of one key intermediate.
One intermediate step involves a hydride transfer from a surrounding NADH molecule to the NO-bound heme (see Appendix~\ref{app:molecule}). The NO-bound heme is expected to exhibit strong correlations and hence can be a good target for quantum computations. The strong correlation arises as $\pi$/$\pi^*$ orbitals of both NO and porphyrin moieties are known to mix wit the 3-d orbitals of the Fe center~\cite{lehnert2021biologically,monsch2019fe,goings2022reliably}.
In this context, we inquire whether electrostatic interactions favour the stabilisation of the NADH molecule in the heme pocket. In practice, this involves calculating the difference in electrostatic energies between both intermediates, denoted as $\Delta E_\mathrm{elst}=E_\mathrm{elst}(\text{B})-E_\mathrm{elst}(\text{A})$.

The preparation of the model systems for the two intermediates of interest, shown in Fig.~\ref{fig:modelAB},
is discussed in great detail in Appendix~\ref{app:molecule}.
To calculate the electrostatic interactions, we break down both intermediates into two separate monomers. For intermediate A, the first monomer consists of the NO-bound heme, while the second monomer encompasses the amino acids and water molecules. In the case of intermediate B, the first monomer includes both the NO-bound heme and the NADH molecule, while the amino acids and the remaining water molecule form the second monomer.
In both cases the ground state wavefunction of the first monomer is partially treated with a quantum computer. 
To match with the hardware constrains, we select a small four electrons in four \glspl{mo} (4e, 4o) active space. 
\begin{figure}
    \centering
    \includegraphics[width=0.9\linewidth]{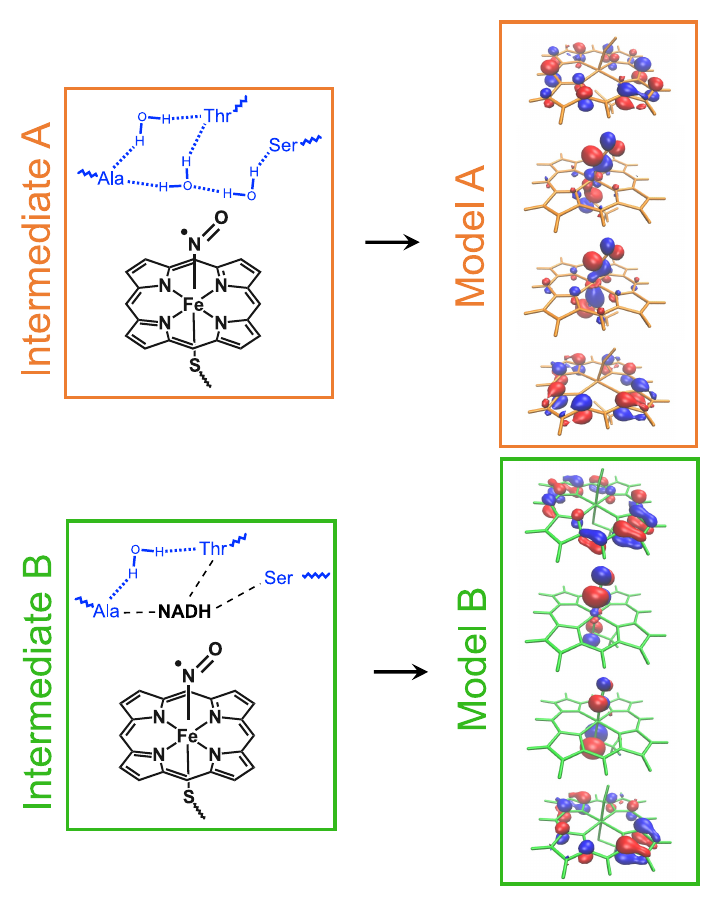}
    \caption{(left) 2D representations of intermediate A and B. They include the NO-bound heme, three surrounding amino acids, water molecules and NADH (if present). In both cases, monomer $A$ and $B$ are shown in black and blue, respectively.  (right) The four orbitals in the active space of monomer $A$ for each intermediate.}
    \label{fig:modelAB}
\end{figure}
Capturing the important electronic correlations in our systems would obviously require a much larger active space~\cite{goings2022reliably} and the choice we make is strictly driven by the experimental setup. 
However, the orbitals selected by looking at the natural orbital occupation numbers (see Appendix~\ref{app:molecule} and Fig.~\ref{fig:modelAB}) include character from the iron 3d, NO (anti)bonding $\pi$, porphyrin  (anti)bonding $\pi$ orbitals which govern the strong corrleation in this system~\cite{lehnert2021biologically,monsch2019fe,goings2022reliably}.

\begin{figure}
\includegraphics[width=0.9\linewidth]{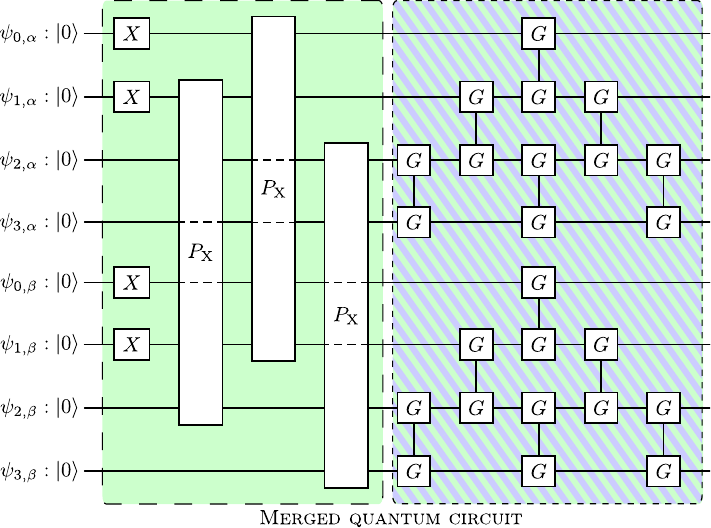}
\caption{The quantum circuit used in this work. The two-qubit gates labeled with $G$ denote Given rotations, which are equivalent to local fermionic basis rotations after a Jordan-Wigner mapping. The four-qubit $P_\mathrm{X}$ gate represents a PairExchange gate \cite{Anselmetti_2021}. We refer to Appendix~\ref{app:quantum_circuit} for more details. The Givens rotations of the \gls{vqe} circuit were merged with the basis rotation circuit (hatched area) to reduce the circuit depth.}
\label{fig:quantum_circuit}
\end{figure}

To find the ground state energy in the active space, the \gls{vqe} algorithm \cite{peruzzo2014variational} is used. The fermionic Hamiltonian in the (4e, 4o) active space is mapped through the Jordan-Wigner mapping onto a 8-qubit system, in $\alpha$-then-$\beta$ ordering, meaning the first (last) 4 qubits are assigned to the $\alpha$/spin-up orbitals ($\beta$/spin-down orbitals) with a qubit in state $\ket{1}$/$\ket{0}$ representing an occupied/unoccupied orbital.
We adapt the \gls{vqe} ansatz from Ref.~\cite{Anselmetti_2021} to generate an approximate ground state. This ansatz is composed out of an alternating application of Givens rotations, $G(\theta)$ (i.e. fermionic basis rotations represented after Jordan-Wigner transformation) and fermionic double excitations, $P_X(\theta)$, where we restrict our ansatz to a single layer of each, see Fig.~\ref{fig:quantum_circuit} and Appendix~\ref{app:quantum_circuit}. The set of parameters yielding an approximation of the ground state is found by optimizing the expectation value of the second quantized Hamiltonian with the L-BFGS-B algorithm using a classical simulator implemented in Qiskit \cite{Qiskit}.  Independent from the number of layers, the last action of the ansatz corresponds to a Givens rotation network. As the electrostatics require only the diagonal part of the \gls{1pdm} $\Tilde{\gamma}_{vv}$ in the electrostatic potential natural orbital basis, we implement the fermionic basis rotation $U_{tv}$ that diagonalizes $\JJ^B_{tt'}$  by another network of Givens rotation gates identical in structure to this last part of the ansatz. 
We leverage the property that the multiplication of two Givens rotation networks is itself another Givens rotation network, see Appendix~\ref{sec:quantum_circuits} for details.
This allows us to merge the two Givens rotation networks at the end of our circuit and reduce the circuit depth, see Fig.~\ref{fig:quantum_circuit}. We finally measure the qubits in their computational basis to obtain the expectation value of the electrostatics operator in the active space and combine it with the contribution from the core as explained in Appendix~\ref{app:active_space_sapt}.

\section{Hardware implementation \& Results}
\label{sec:results}

As quantum hardware, we use the quantum processor \textit{aqt\_marmot} hosted by AQT \cite{aqt}. The \textit{aqt\_marmot} system is based on trapped $^{40}$Ca$^+$ ions and supports a universal set of gates. The native gate set comprises single-qubit gates with arbitrary rotation angles and axis. The entangling operation is a two-qubit M\o lmer-S\o rensen (MS) gate \cite{sorensen2000entanglement}. The MS gate allows for entangling operations with arbitrary rotation angles that can be implemented between any qubit pair (see Appendix~\ref{app:quantum_circuit}). In this work, we utilize an 8-qubit register featuring all-to-all connectivity. 
We choose $R_\mathrm{X}(\theta)=\mathrm{exp}[-\mathrm{i}\theta/2 \hat{X}_{i}]$, $R_\mathrm{Z}(\theta)=\mathrm{exp}[-\mathrm{i}\theta/2 \hat{Z}_{i}]$ and $R_\mathrm{XX}(\theta)=\mathrm{exp}[-i\theta \hat{X}_{i}\hat{X}_{j}]$ with $ \hat{X}_{i}$ and $ \hat{Z}_{i}$ being the Pauli-X and Pauli-Z matrices acting on qubit $i$ respectively, as the basis gate set since it closely resembles the native gate set of the \textit{aqt\_marmot} system. The gate errors can be approximated by depolarizing noise acting on the addressed qubits. The error rates for the single-qubit gates ($R_\mathrm{X}$ error) are approximately $3\cdot10^{-4}$ on average, whereas the error rate for the two-qubit gate ($R_\mathrm{XX}$ error) is around $1.5\cdot10^{-2}$ on average. The typical gate times are $15\,\mu$s for single-qubit gates and $200\,\mu$s for two-qubit gates. The T$_1$ and T$_2$ times are $1.14 \pm 0.06$ seconds and $0.452 \pm 0.068$ seconds respectively, resulting in a coherence/gate time ratio of about $10^3$ \cite{aqt_coherence}. 

We use the quantum circuit described in Section~\ref{sec:classical} to prepare an approximate ground state on the quantum computer. We compile the quantum circuits' gates into the hardware's native gate set to implement the circuit on the trapped-ion quantum computer. Subsequently, we use Qiskit's transpiler~\cite{Qiskit} to further optimize the circuit and minimize the total number of gates. After optimization, our circuit consists of 279 single-qubit gates and 63 two-qubit gates. 
The transpiled circuit for intermediate A is displayed in Appendix~\ref{app:quantum_circuit}.

\begin{figure*}
\centering
\includegraphics[width=0.9\linewidth]{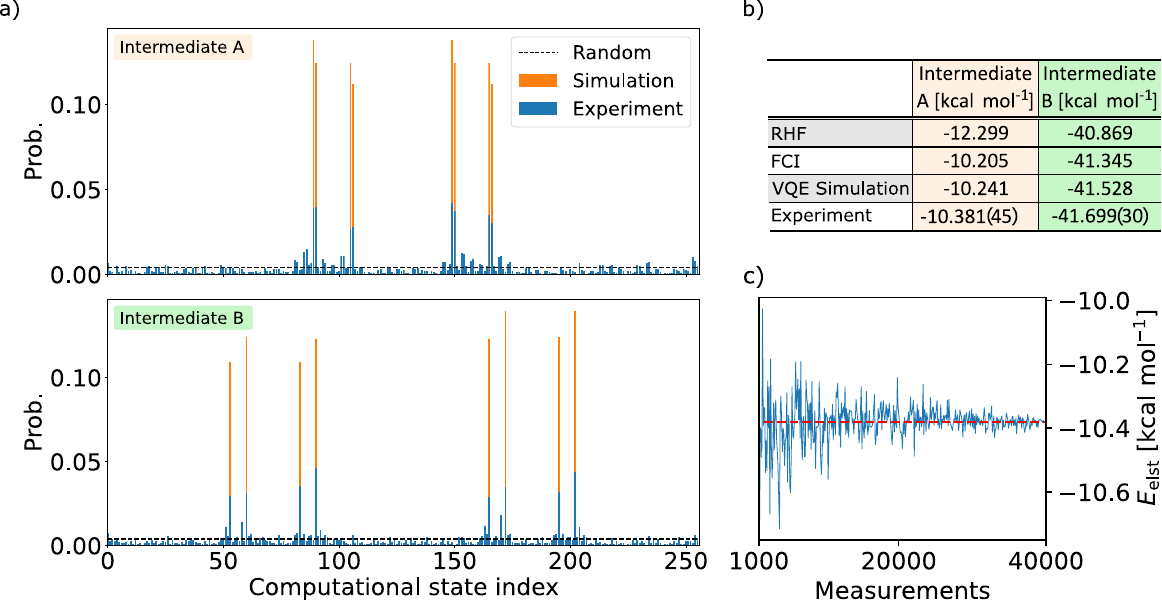}
\caption{(a) The output statistics of the \gls{vqe} circuit, including the final basis rotation for intermediate A (top) and intermediate B (bottom). The experimental outcome (blue) is compared against exact classical simulation results (orange) and random sampling (black line). For each experiment, $N_\mathrm{meas.}=4\times 10^4$ measurements were used. (b) A comparison between the electrostatic energies obtained from various classical and quantum methods. (c) The convergence of the electrostatics of intermediate A as a function of the number of measurements used to construct the \gls{1pdm} in Eq.~(\ref{eq:diag_meas}).}
\label{fig:experimental_results}
\end{figure*}
In Figure~\ref{fig:experimental_results}(a) we present the output statistics obtained from $N_\mathrm{meas.}=4\times 10^4$ measurements for intermediate A and B. We compare the experimental results to results from an error-free \gls{vqe} simulation on a classical computer. 
We also put these distributions in contrast with the uniform one (random sampling).
Despite a non-negligible level of noise, we observe strong agreement in the dominant computational basis states.
To obtain the electrostatic interaction energies for both intermediates, we first exclude all computational states with incorrect alpha and beta particle numbers from the measured data, technique commonly referred to as \textit{postselection}. In both experiments, we discard roughly 50\% of the samples. We then construct the diagonal parts of the rotated \gls{1pdm} using Eq.~(\ref{eq:diag_meas}) and compute the electrostatics according to Eq.~(\ref{eq:electrostatics_from_rotated_rdm}). We summarize our findings in Fig.~\ref{fig:experimental_results}(b). We find that the electrostatic energies of both systems are within chemical accuracy ($1$ kcal mol$^{-1}$) to the error-free VQE simulations, with deviations of $\Delta E=-0.204(45)$ kcal mol$^{-1}$ and $\Delta E=-0.354(30)$ kcal mol$^{-1}$ for intermediate A and intermediate B, respectively. To estimate the statistical errors, we calculate the electrostatics energy for each measurement with correct particle number separately. From the list of electrostatics, we determine the standard error of the mean, given by $\sigma_\mathrm{E}=\sigma/N_\mathrm{cor.}$, where $\sigma$ is the standard deviation of the list and $N_\mathrm{cor.}$ the number of measurements with correct particle number. In Fig.~\ref{fig:experimental_results}(c), we also report the convergence of the electrostatics when increasing the number of measurements to construct the required \gls{1pdm}. Here, for clarity, we report estimates obtained from using more than $1000$ measurements. 

In Appendix~\ref{app:additional_experiments} we show the same convergence results using $2$ to $1000$ measurements. In this case, the calculated electrostatic energies lie in a much broader range (from $-26.98$ kcal mol$^{-1}$ to $6.80$ kcal mol$^{-1}$). This emphasizes that the spectrum of our estimate is non-trivial but that its expectation value converges rapidly when increasing the number of measurements.
This also indicates that the remaining $\Delta E$ is a consequence of device noise biasing our estimate. In Appendix~\ref{app:additional_experiments}, we also explore results from random sampling, which yields fairly accurate electrostatic energies. By comparing the output statistics between experimental results, noise-free simulations, and random sampling, we argue that the observed phenomenon is an attribute of the system under investigation and does not undermine the experimental outcome.

From the experimental results, we calculate a difference in electrostatic energies of $\Delta E_\mathrm{elst}^\mathrm{exp.}=-31.32(5)$ kcal mol$^{-1}$ between intermediate A and B, allowing us to answer the question raised in Sec.~\ref{app:molecule} and conclude that the electrostatics favors the stabilization of NADH in the heme pocket. Although, for this model system, this could have been qualitatively described using Hartree-Fock theory ($\Delta E_\mathrm{elst}^\mathrm{RHF}=-28.57$ kcal mol$^{-1}$), our results quantitatively align with FCI calculations ($\Delta E_\mathrm{elst}^\mathrm{FCI}=-31.14$ kcal mol$^{-1}$).

Multiple efforts to reduce the noise bias $\Delta E$ were explored: To investigate and mitigate potential biases stemming from keeping the ion-qubit assignment fixed during the quantum computation, we performed additional experiments, where we introduced a random alteration of the qubit-ion assignment every 100 experimental runs. Additionally, to mitigate the impact of qubit decay from $\ket{1}\rightarrow\ket{0}$, we treated the qubits' zero state $\ket{0}$ to represent an occupied orbital, while $\ket{1}$ represent an unoccupied orbital. This can be easily implemented by adding a full layer of Pauli-X gates in the beginning of the quantum circuit while concurrently substituting all circuit parameters with their negated values. 
As demonstrated in Appendix~\ref{app:additional_experiments}, incorporating these error mitigation techniques does not significantly improve the accuracy of the electrostatic energy. We also classically simulated the effect of Zero-Noise-Extrapolation (ZNE), an error mitigation method well-suited for depolarizing errors as they predominate in the used quantum processor (see Appendix~\ref{app:zne-extrapolation}). While we found an improvement of the electrostatic energy, the error of the estimate increased significantly. To reduce the error, a substantial increase in the number of experimental measurements would have been necessary.

In order to draw a comparison between our approach and the supermolecular approach given in Eq.~(\ref{eq:interaction_energies}), we would require to estimate the ground state energies of both monomers and the larger dimer system. However, since we only measure the diagonal elements of the rotated \gls{1pdm} of monomer $A$, as per Eq.~(\ref{eq:diag_meas}), we cannot use these measurements to estimate the ground state energy of monomer $A$, as this would necessitate the full \gls{1pdm} and the \gls{2pdm}. Instead, we use these measurements to compute the expectation value of the diagonal elements of the one-body part of the Hamiltonian, with further details provided in App.~\ref{app:one_body}. An deviation of $1.16$ kcal mol$^{-1}$ was observed when comparing the experimental results with the noise-less simulation. Considering the error in the diagonal contribution to the one body energy, it is unlikely that the accuracy of the ground state energies of the monomer systems would be sufficient to estimate interaction energies within chemical accuracy, let alone the ground state energy of the larger dimer system.

\section{Conclusion}

In our study, we performed the first direct quantum computations of interaction energies; specifically we computed the electrostatic energies of two important intermediates in the catalytic conversion of \ce{NO} to \ce{N2O} by P450nor. Using a \gls{vqe} ansatz, we generated an approximate ground state for one of both monomers on a trapped-ion quantum computer, while the other was treated with classical \gls{hf} theory. A fermionic basis rotation at the end of the quantum computation allowed the efficient  measurement of the electrostatics energy with a single quantum circuit and without extending the circuit length. We found that the quantum-computed electrostatics strongly agree with our simulation results, within a chemical accuracy of $1$ kcal mol$^{-1}$. As measuring under these fermionic basis rotations is also used when e.g. measuring ground state energies efficiently with double factorization \cite{oumarou2023accelerating, hohenstein2023efficient, von2021quantum}, this experiment implies that these techniques are starting to be in reach of current hardware.

In order to execute the demonstrated quantum experiment, we had to considerably simplify the computational problem to conform to limited number of qubits, coherence times and gate fidelities available on the quantum computer.  
Furthermore, we employed classical simulations of the quantum circuit to optimize the variational parameters. This allowed us to leave aside the task of finding the ground state in a noisy environment and focus our study solely on the computation of the electrostatic energy. 
While this work does not offer new insights into the systems' chemistry, it does provide the first experimental demonstration of the potential usefulness of quantum algorithms which are tailored to the computation of quantities beyond ground state energies.
In particular, we showed how the measurement count can be reduced in practice, resulting, for the small models considered, in a shot noise resilient estimate.

Scaling our approach to industry-relevant problems requires further research. First, the electrostatic energy is only a single term of \gls{sapt}, which by itself does not require symmetry adaption. Obtaining other terms such as the exchange, induction and dispersion is vital to accurately describe the interaction between two molecules. To this end, efficient measurement strategies similar to this work must also be developed. 
Another crucial aspect for future research resides in finding efficient measurement schemes when both monomers exhibit strong correlations and require accurate descriptions on quantum computers.  

While tailored quantum computations, such as the one presented, allow to push the boundaries of existing quantum hardware, it is unlikely that a quantum advantage will become possible without exploiting any form of error mitigation or limited error correction. Fault-tolerant quantum computations of \gls{sapt} energies~\cite{cortes2023fault} could potentially offer another approach to the quantum computation of interaction energies.\\

\section*{acknowledgements}
We thank Clemens Utschig-Utschig for insightful discussions. We thank Edward G. Hohenstein for his valuable feedback on this manuscript. 
AQT acknowledges support by 
the European Innovation Council under the QCDC Grant Agreement no. 190118992,
the European Union’s Horizon Europe research and innovation program under grant agreement
number 101114305 (“MILLENION-SGA1” EU Project), 
the Austrian Research Promotion Agency under contract number 897481 (HPQC),
884471 (ELQO),
and the Austrian Science Fund under project number F7109 (SFB BeyondC).
R.M.P., M.L. and P.J.O. own stock/options in QC Ware Corp.
\subsection*{Hardware and data availability}
The presented measurements were performed on the full-stack quantum computer \textit{aqt\_marmot}.
We would like to point out that this particular device is no longer online and available, but it was replaced with a new system hosted by AQT.
The measured data is available from the corresponding author upon reasonable request.
\bibliographystyle{apsrev4-2}
\bibliography{biblio}

\clearpage
\onecolumngrid
\appendix

\section{Expression of the electrostatics in terms of the 1-PDM}
\label{app:Eelst}

In this section we formally define the electrostatic energy in terms of the spin-summed \gls{1pdm} which, in spatial coordinates is, 
\begin{equation}
    \gamma(\vec{r}_1, \vec{r}_{1'}) \equiv \int \td \sigma_1 \, \int \td \sigma_{1'} \,\int \td \vec{x}_2 \,... \int \td \vec{x}_N \,\, \Psi(\vec{x}_1,...,\vec{x}_N)\Psi(\vec{x}_{1},...,\vec{x}_{N}) 
    \equiv \sum_{pp'} \gamma_{pp'}\phi_p(\vec{r}_1)\phi_{p'}(\vec{r}_{1'})
\end{equation}
where $\gamma_{pp'}$ is the spin-summed \gls{1pdm} in a (non-orthogonal) spin-restricted spatial orbital basis $\{\phi_p(\vec{r})\}$.

In this notation, the electrostatic energy can be written as
\begin{equation}
    E_{\text{elst}} = \sum_{pp'} \sum_{qq'} \gamma^A_{pp'} (pp'|qq') \gamma_{qq'}^B + \sum_{pp'}\gamma_{pp'}^A V_{pp'}^B + \sum_{qq'}V^A_{qq'}\gamma_{qq'}^B + V_{AB}.
    \label{eq:Eelst_si}
\end{equation}
Here, $\gamma_{pp'}^A$ is the \gls{1pdm} of monomer $A$. 
The two electron integrals are defined as
\begin{equation}
    (pp'|qq') = \int_{\R^6} \td^3\vec{r}_1 \td^3\vec{r}_2 \,\, \phi_p(\vec{r}_1) \phi_{p'}(\vec{r}_1) \frac{1}{r_{12}} \phi_q(\vec{r}_2) \phi_{q'}(\vec{r}_2).
\end{equation}
The nuclear potential, $V_{pp'}^A$ of monomer $A$ is
\begin{equation}
    V_{pp'}^A = - \sum_{\mathcal{A}}^{M_A} \int_{\R^3} \td \vec{r}_1 \,\, \phi_p(\vec{r}_1)\phi_{p'}(\vec{r}_1)\frac{Z_\mathcal{A}}{r_{1\mathcal{A}}}
\end{equation}
where $M_A$ is the number of atoms in monomer $A$ and $Z_{\mathcal{A}}$ is the nuclear charge of atom $\mathcal{A}$.
Finally, the inter-monomer nuclear repulsion is given by
\begin{equation}
    V_{AB} = \sum_{\mathcal{A}}^{M_A}\sum_{\mathcal{B}}^{M_B}\frac{Z_{\mathcal{A}}Z_{\mathcal{B}}}{r_{\mathcal{AB}}}.
\end{equation}

Eq.~(\ref{eq:Eelst_si}) and Eq.~(\ref{eq:Eelst_main}), in the main text, are easily related by noting that the number of electrons in monomer $A$ is
\begin{equation}
    N_A = \sum_{pp'} S_{pp'}\gamma_{pp'}^A
\end{equation}
where $S_{pp'}$ is the overlap matrix,
\begin{equation}
    S_{pp'}\equiv \int_{\R^3} \td^3 \vec{r}_1 \,\, \phi_p(\vec{r}_1)\phi_{p'}(\vec{r}_1).
\end{equation}

\section{Preparation of intermediates A and B}
\label{app:molecule}

\begin{figure}
    \centering
    \includegraphics[width=0.5\linewidth]{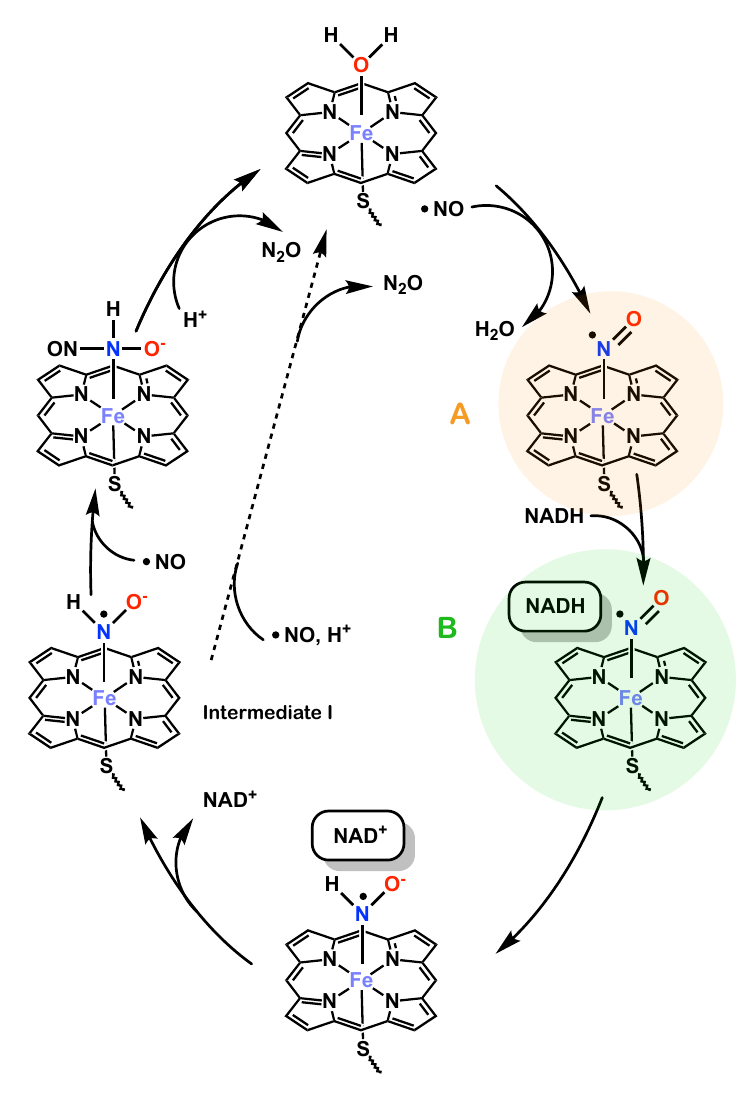}
    \caption{Reaction scheme of the reduction of NO to N$_2$O catalysed by P450nor. In our study we focus on \ce{NO}-bound intermediates A and B and the question whether the electrostatic interaction favour the stabilisation of the NADH molecule in the heme pocket.}
    \label{fig:cat-cycle}
\end{figure}

\begin{figure}
    \centering
    \includegraphics[width = 0.7\linewidth]{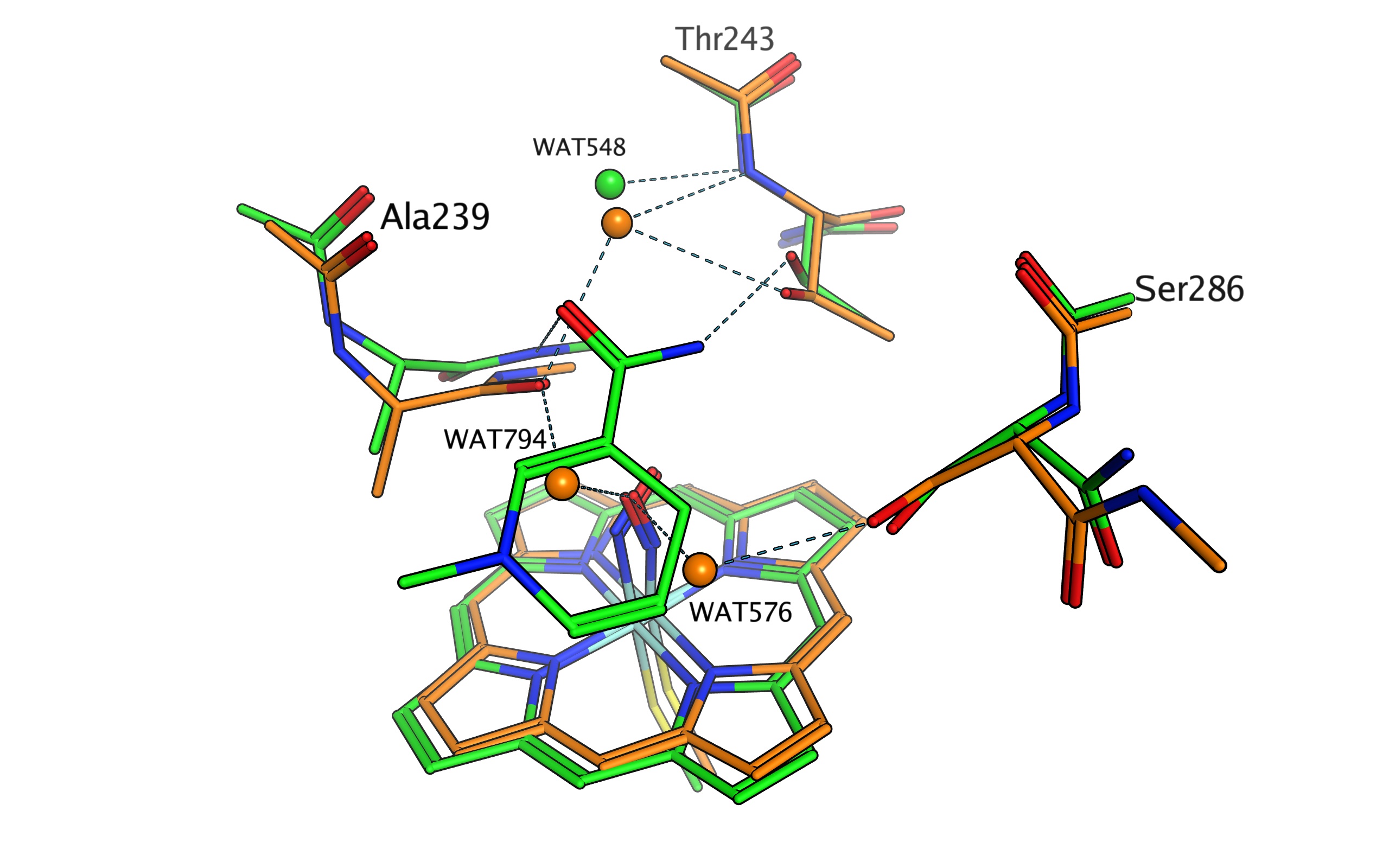}
    \caption{Overlay of the model complexes before DFT optimization. Intermediate A and B are shown in orange and green, respectively. Dashed lines denote hydrogen bonds and the water labels correspond to pdb entry 1CL6.}
    \label{fig:structures_before_dft}
\end{figure}

We extract models for intermediates A and B from highly resolved crystal structures taken from the \gls{pdb}. The \gls{pdb} entries are 1CL6~\cite{shimizu2000proton} and 1XQD~\cite{oshima2004structural} for intermediate A and B, respectively.
Both structures are first subjected to the Protein Structure Preparation Procedure as implemented in MOE~\cite{moe}, which adjusts hydrogens, partial charges, non-modeled side chains, and hybridizations. 
Our focus is on the active site. Therefore we prepare model systems that contain most of the important structural features around the heme binding site. 
The heme iron is coordinated to Cys352 from below, and we truncate this ligand to a methylthiolate ion to resemble the electronic effects of cysteine on the iron. All side chains are removed from the heme itself, as they do not interfere with the binding site. 
On the NO-binding site, we keep the three amino acids Ala239, Thr243, and Ser286 as these residues are strongly involved in a hydrogen bond network including waters and NADH (if present). The three amino acids are kept in their respective orientations from the \gls{pdb} files, the amide bonds along the protein backbone are cut, and the rest of the protein is removed. 
To maintain the correct chemical environment of the three amino acids, the termini are capped to resemble amide bonds (acetylation of the N-terminus and amidation of the C-terminus), thereby ensuring the correct amide bond orientation according to the X-ray structure.
In the model of intermediate A, the NO ligand bound to the heme and additional three water molecules (WAT548, WAT576, WAT794 in 1CL6, as displayed in Figure~\ref{fig:structures_before_dft}) are included as they form a strong hydrogen bond network within the three amino acids and the NO ligand. 
In the model of intermediate B, additional steps are required to model the active site. 
A water (WAT565) from the X-ray structure, hydrogen-bonded to Thr243 is included. 
The carboxylic acid group is modified to a carboxamide group (as in NADH) so that the carbonyl group still forms a hydrogen bond to the backbone NH from Ala239. 
The pyridine ring is reduced by adding a hydrogen atom (to have the reduced form: NADH), and the ligand is capped after the first ribose-carbon. The rest of the NADH is assumed not to contribute to the hydride-transfer reaction. 
The NO-ligand is manually put to the heme to finalize the model system with all relevant and correct species. 
Then, a force field optimization of the hydrogen placements in MOE (AMBER10:EHT) is done, followed by a tethered minimization (tether on all heavy atoms).
An overlay of the two resulting model structures is shown in Fig.~\ref{fig:structures_before_dft}. 

\begin{figure*}
    \centering
    \includegraphics[width=0.9\linewidth]{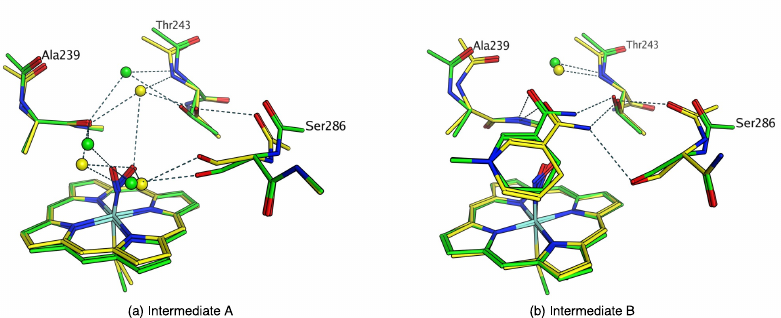}
    \caption{Overlay of the structures of both intermediates before (green) and after (yellow) geometry optimization via DFT. Dashed lines denote hydrogen bonds.}
\label{fig:dft_geometries}
\end{figure*}

Finally, both structures are optimized with \gls{dft}. The position of the heme is held fixed and the position of the C$_{\alpha}$ atoms of the amino acids is also constrained as a surrogate for the protein backbone environment that keeps amino acid side chains and co-factors in place.  
We use the GPU accelerated \gls{dft} implementation of Promethium~\cite{promethium} on a single A100 GPU, allowing us to perform the full geometry optimization at B3LYP-D3/def2-SV(P) level of theory in 3h58 for intermediate A (1042 basis functions) and 6h36 for intermediate B (1130 basis functions). The resulting geometries are shown in Fig.~\ref{fig:dft_geometries}.

\begin{figure}
\centering
    \includegraphics[width=0.49\columnwidth]{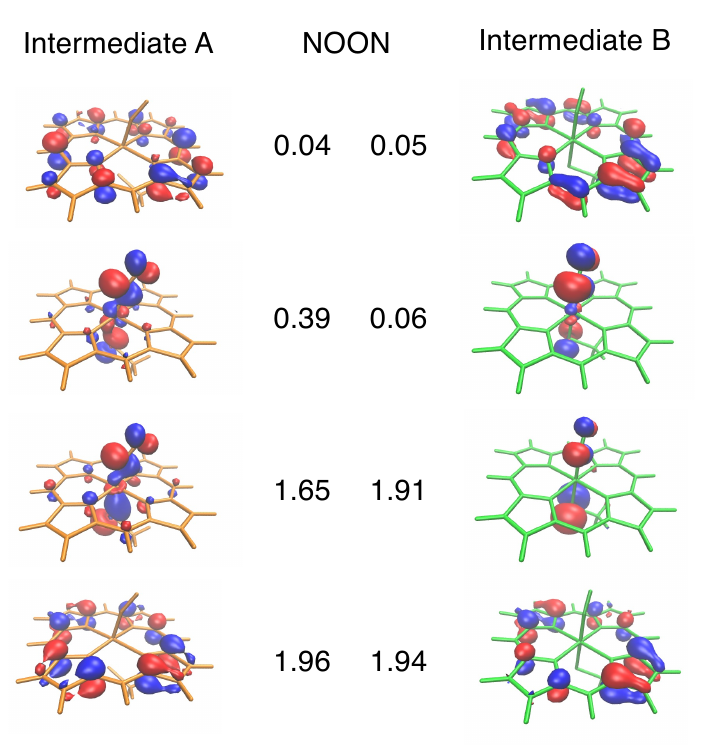}
\hfill
    \includegraphics[width=0.49\textwidth]{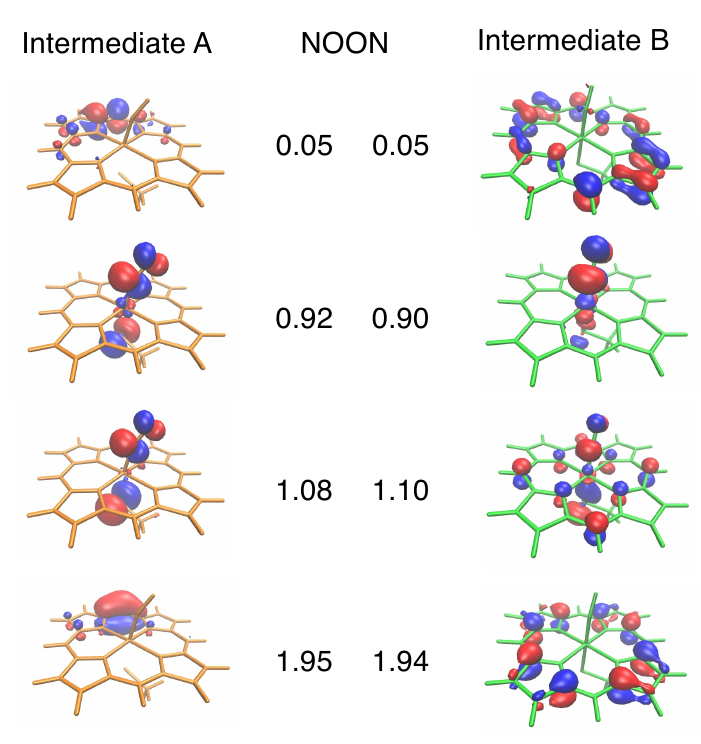}
\caption{(left) The four \gls{shci} natural orbitals around the gap and their occupation numbers, (right) The four \gls{casscf} molecular orbitals around the gap and their occupation numbers}
\label{fig:as}
\end{figure}

Because we are limited quantum resource available on current hardware, we devise an active space of four \glspl{mo}. The size of this active space is solely chosen to fit the hardware constraints and does not reflect the important chemistry in our systems. Therefore, we choose to employ an automated active space selection approach rather than picking the orbitals fully based on their symmetries. 
To do so, we first run a \gls{shci} (as implemented in Dice~\cite{sharma2017semistochastic, holmes2016heat}) for the two systems in the singlet state with $\epsilon = 0.01$, the def2-SVP basis set, in a 20 electrons in 20 orbitals (taken around the HOMO-LUMO gap). This starting active space is sizeable enough to include the iron 3d, iron 4s, iron-nitrogen anti-bonding and axial ligand (anti-)bonding orbitals~\cite{goings2022reliably}. 
From the resulting \gls{1pdm}, we obtain the natural orbitals and order them according to their occupation numbers. We show the four orbitals around the gap in Fig.~\ref{fig:as}(left) along with their occupation numbers. 
These orbitals are a mixture of the iron 3d, the iron-nitrosil (anti)bonding $\pi$, the sulfur p as well as heme $\pi$ orbitals which are all important to capture the essential mixing.
Hence, as a final step we run a \gls{casscf} calculation with PySCF~\cite{sun2018pyscf} using the \gls{shci} natural orbital as a starting guess and a 4 electrons in 4 orbitals active space (4e, 4o).
The resulting orbitals define our final (4e, 4o) active space. They are displayed in Fig.~\ref{fig:as}(right). In both intermediates, two orbitals show strong deviation from integer values indicating the multi-reference character of the systems and exhibit the expected metal ligand backbonding of Fe and NO and porphyrin $\pi$ and $\pi^*$ orbitals. Note that the inclusion of the first and last orbitals in the active space (the one with occupation numbers close to 2 and 0, respectively) lowers the total energy by 18 mHa for intermediate A and 10 mHa for intermediate B, justifying their importance in the ground state.

\section{Active space description of electrostatics}
\label{app:active_space_sapt}
In the following, we describe how to recover the electrostatic energy in the case where the quantum computer is used to generate an approximate ground state in an active space.

Since the \gls{1pdm} is block diagonal, we can rewrite Eq.~\ref{eq:Eelst_mo} as
\begin{equation}
    E_{\text{elst}} = \sum_{t,t'=0}^{N_\text{core}} \JJ^B_{tt'} \gamma_{tt'}^A + \sum_{t,t'=N_\text{core}}^{N_\text{act}}\JJ^B_{tt'} \gamma_{tt'}^A
\end{equation}
where in the first (second) term the indices run over the \glspl{mo} in the core (active) space. 
In the core space, by definition $\gamma_{tt'}^A$ is diagonal (with $2$s on its diagonal).
Hence, we only rotate the orbital basis in the active space and, as per Eq.~(\ref{eq:electrostatics_from_rotated_rdm}), we find
\begin{equation}
    E_{\text{elst}} = 2\sum_{t=0}^{N_\text{core}} \JJ^B_{tt} + \sum_{v=N_\text{core}}^{N_\text{act}}\bar{w}_v\bar{\gamma}^A_{vv}.
\end{equation}

\section{Quantum circuits}
\label{sec:quantum_circuits}
\begin{figure*}
\includegraphics[width=0.9\linewidth]{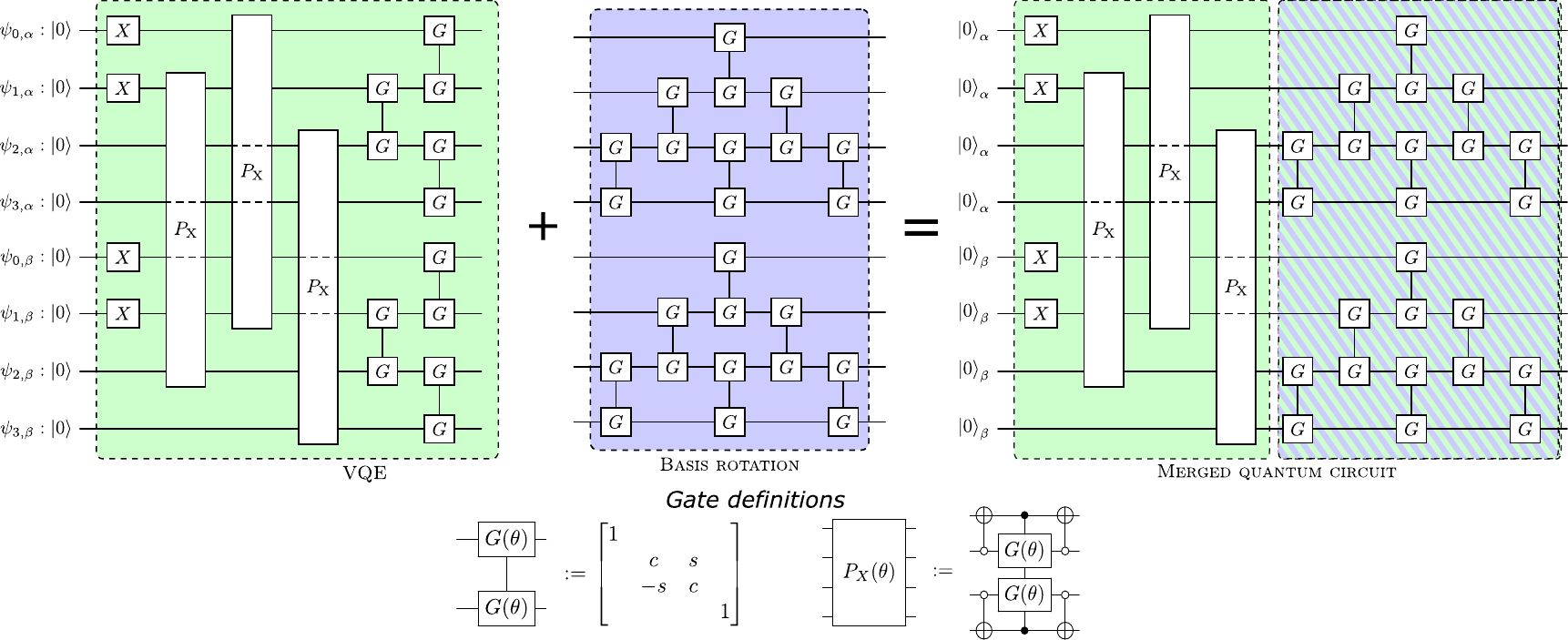}
\caption{(top) The quantum circuit used in this work. The two-qubit gates labeled with $G$ denote Given rotations, which are equivalent to local fermionic basis rotations after a Jordan-Wigner mapping. The four-qubit $P_\mathrm{X}$ gate represents a PairExchange gate, see below. The green circuits represent the \gls{vqe} ansatz, while the blue circuit represents the basis rotation to efficiently measure the electrostatic interaction. On the right hand side, the Givens rotations network of the \gls{vqe} circuit was merged with the basis rotation circuit (hatched area). (bottom) Gate definitions of the Givens rotation gate $G(\theta)$ and the PairExchange Gate $P_X(\theta)$ as introduced in \cite{Anselmetti_2021}. Matrix entries are $c := \cos (\theta/2), s := \sin (\theta/2)$.}
\label{fig:quantum_circuit_with_rotation}
\end{figure*}
\label{app:quantum_circuit}

The \gls{vqe} circuit was designed on the model of the quantum number preserving ansatz of Ref.~\cite{Anselmetti_2021}. 
This ansatz is composed out of an alternating application of Givens rotations, $G(\theta)$ (i.e. fermionic basis rotations represented after Jordan-Wigner transformation) and fermionic double excitations, $P_X(\theta)$.
To maximally reduce the circuit depth while keeping a general architecture, ignoring a priori knowledge of the targeted ground state, we follow three steps. 
First, we keep only one layer of each kind ($G(\theta)$ and $P_X(\theta)$). 
Second, in both layer kinds, we start by entangling the last occupied and first unoccupied qubits (as defined in the initial \gls{hf} state). We then progressively entangle the next two qubits up and down, in a ladder way, until we reach to first and last qubit, respectively. In our small 4+4 qubit system this only translates to 3 $P_X(\theta)$ and 3+3 $G(\theta)$ gates. Third, we first add the $P_X(\theta)$ layer and then the $G(\theta)$ layer.
This setup allows us to merge the $G(\theta)$ layer with the subsequent Givens rotation network, representing the fermionic basis rotation into the electrostatic potential natural orbital basis, shown in Sec.~\ref{sec:sapt_elec}.
We use the fact that two Givens rotation network can be easily merged. As every Givens rotation network represents a fermionic basis rotation $U_{tv}$, two Givens rotation networks can be easily merged by multiplying the underlying fermionic basis rotations. From the found new basis rotation, we generate a new Givens rotation network. The circuit architecture together with the merging of the Givens rotation networks is more clearly depicted on Fig.~\ref{fig:quantum_circuit_with_rotation}.

As explained in the main text, we choose the gates $R_\mathrm{X}(\theta), R_\mathrm{Z}(\theta), R_\mathrm{XX}(\theta)$ as target gate set for the Qiskit transpiler, where the $R_\mathrm{XX}(\theta)$ entangling operation is a two-qubit M\o lmer-S\o rensen (MS) gate that can also be described as,
\begin{align}
    U_{\mathrm{MS}}(\theta)=
    \begin{pmatrix}
    c & 0 & 0 & -\mathrm{i}s \\
    0 & c & -\mathrm{i}s & 0 \\
    0 & -\mathrm{i}s & c & 0 \\
    -\mathrm{i}s & 0 & 0 & c \\
    \end{pmatrix}\,,
\end{align}
with $c=\cos{(\theta/2)}$ and s=$\sin{(\theta/2)}$.

\begin{figure*}
    \centering
    \includegraphics[width=0.7\textwidth]{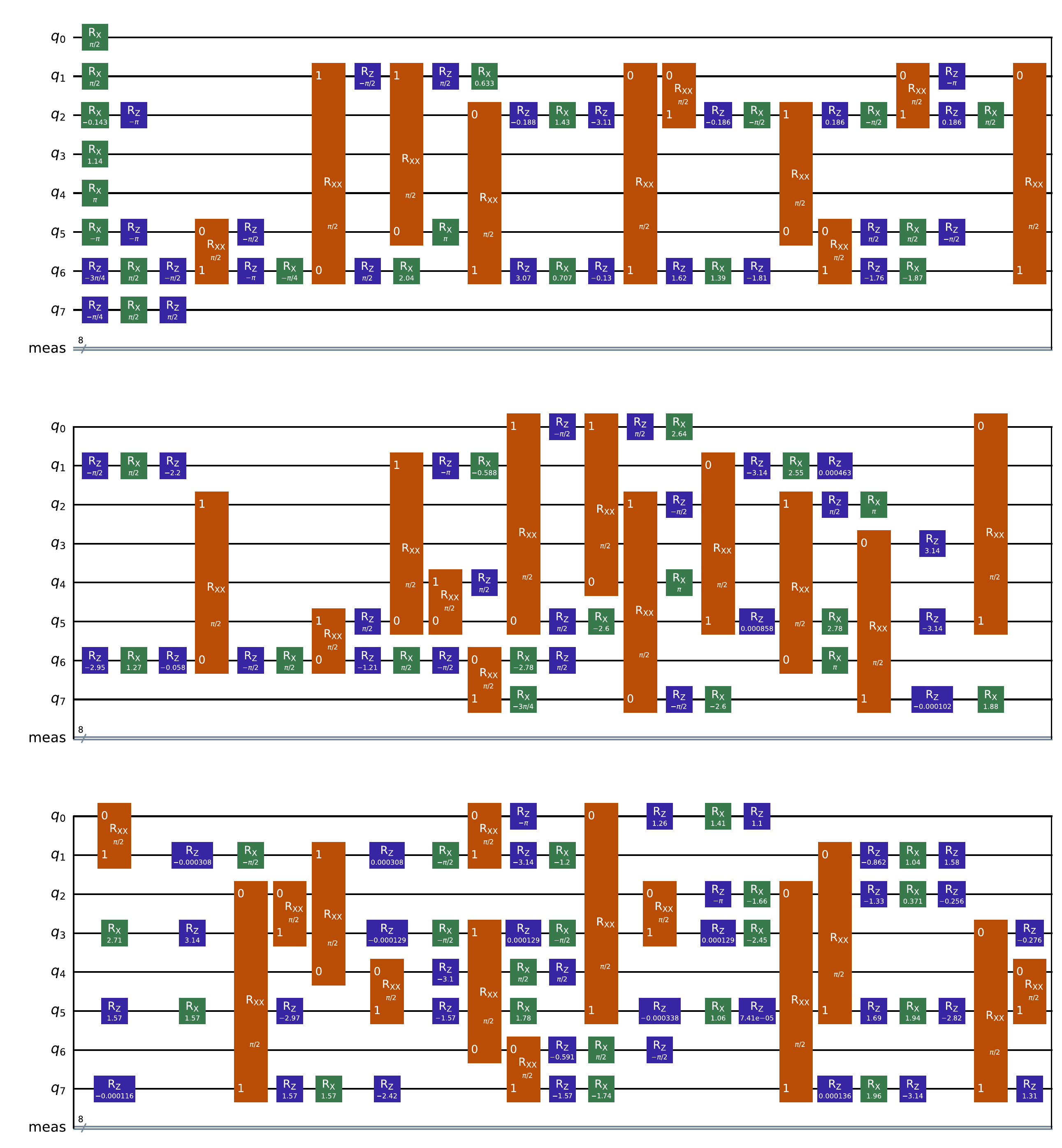}
    \caption{Part 1 of transpiled quantum circuit for intermediate A. Rx, Rz and Rxx gates are shown in green, blue and orange, respectively.}
    \label{fig:intermediate_A_circuit-1}
\end{figure*}

\begin{figure*}
    \centering
    \includegraphics[width=0.7\textwidth]{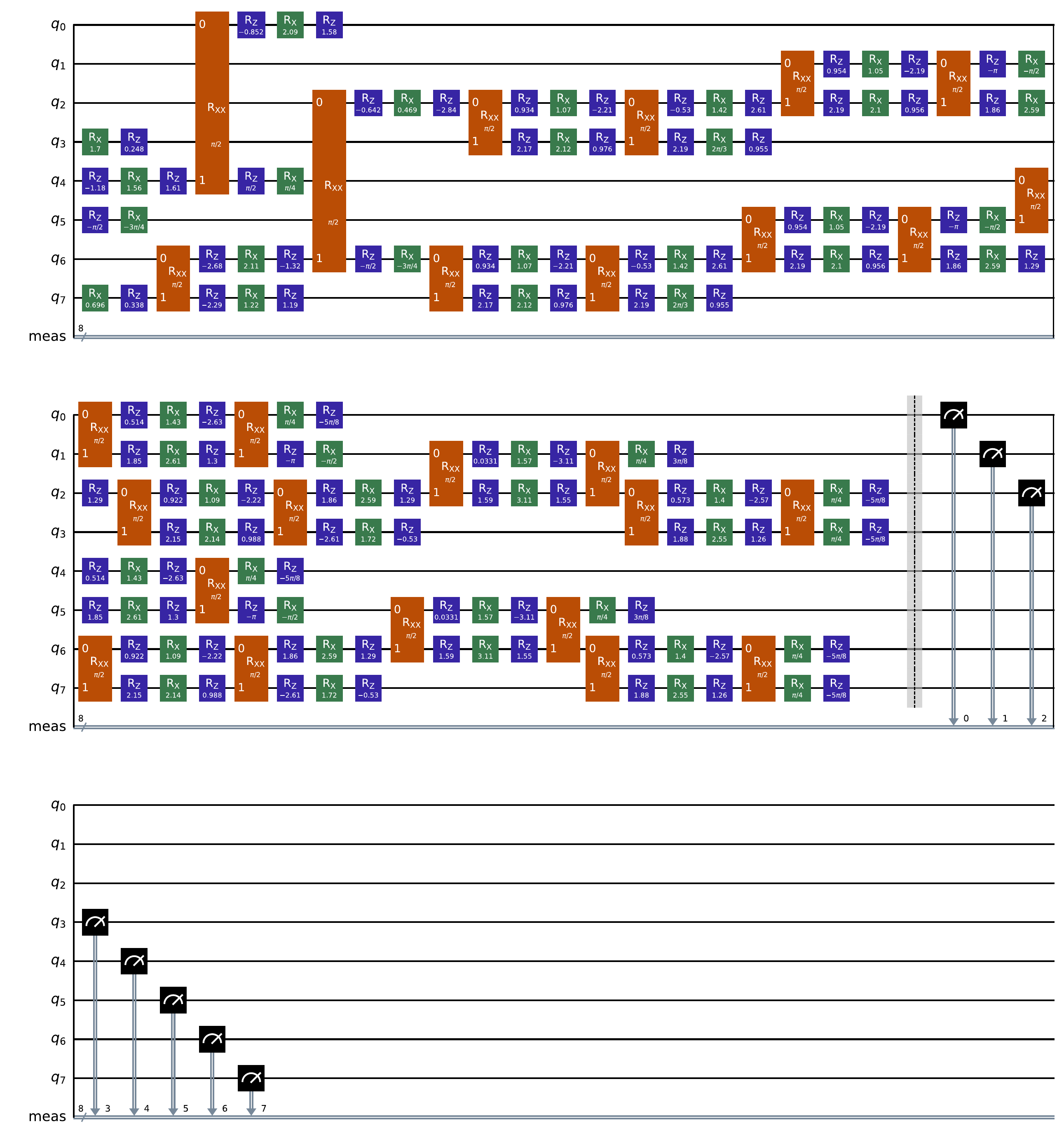}
    \caption{Part 2 of transpiled quantum circuit for intermediate A. Rx, Rz and Rxx gates are shown in green, blue and orange, respectively.}
    \label{fig:intermediate_A_circuit-2}
\end{figure*}

In Figure~\ref{fig:intermediate_A_circuit-1} and Figure~\ref{fig:intermediate_A_circuit-2} we show the full transpiled quantum circuit used to compute the electrostatic energy in intermediate A. The circuit comprises a total number of gates of $(R_\mathrm{Z}, 168), (R_\mathrm{X}, 111), (R_\mathrm{XX}, 63)$. The circuit for intermediate B is similar to the circuit for intermediate A with a total number of gates of $(R_\mathrm{Z}, 168), (R_\mathrm{X}, 113), (R_\mathrm{XX}, 63)$.

Note that for both systems, the optimized angles of the last two $P_X(\theta)$ gates are small ($\approx 10^{-3}$). Their contribution to the energy is therefore negligible here. However, the standard Qiskit transpiler does not do circuit approximation, which is why we implement every circuit unitary as specified. In principle, the circuit size could be further minimized using approximation methods, where the level of approximation would depend on the noise of the specific hardware. While it seems trivial for a 8 qubit circuit with few gates, this emphasizes the need for robust transpilation codes or for guided circuit construction techniques~\cite{grimsley2019adaptive} when scaling up to relevant size systems. 
Here, in the pursuit of reporting the performance of state of the art quantum hardware and software kit, we keep the resulting transpiled circuit while bearing in mind that the noise introduced by applying the previously mentioned $P_X(\theta)$ gates will probably be greater than their contribution to the energy. 


\section{Additional experimental results}
\label{app:additional_experiments}

The electrostatic energy of our model system, constrained to a small active space to align with experimental constraints, exhibits minimal variation in response to state errors. To validate the reliability of our results as a consequence of successful quantum computation rather than mere chance, we present additional experimental data below.
In Table~\ref{tab:additional_experiments}, we first compare the electrostatic energies obtained in classical simulation of an exact superposition state (random sampling) and of the \gls{vqe} state. 
The difference between the two intermediate, $\Delta E_\mathrm{elst} = E_{\text{elst}}(\text{B}) - E_{\text{elst}}(\text{A})$, is -31.751 kcal mol$^{-1}$ for random sampling, -31.314 kcal mol$^{-1}$ for the exact VQE. These values are to be compared to the -31.140 kcal mol$^{-1}$ predicted by FCI. While the VQE results bring us closer to FCI, the random sampling outcomes also fall within chemical accuracy, making it challenging to definitively conclude the true success of the experiment from looking exclusively at the electrostatic energy.
For this reason, in Table~\ref{tab:overlap_experiements}, we illustrate the correspondence between the statistics derived from the noise-less simulation of the quantum circuit and both the experimental results and the exact superposition distribution associated with random sampling. Here, the consistency of the results becomes more evident, with the overlaps of the experimental distributions significantly surpassing those obtained from random sampling.

For each intermediate, we also present 4 different experiments: two solely derived from the VQE quantum circuit (plain), and the remaining two incorporating error mitigation techniques (mitigated). 
In particular, we aim to mitigate potential biases coming from the difference in qubit performance and from qubit relaxation ($\ket{1}\rightarrow\ket{0}$).
Therefore, in these additional experiments, we introduced a random alteration of the qubit-ion assignment every 100 experimental runs along with an inversion of the qubits' $\ket{0}$ and $\ket{1}$ states to represent an occupied orbital and an unoccupied orbital respectively. 
This last bit can be easily implemented by adding a full layer of Pauli-X gates in the beginning of the quantum circuit while concurrently substituting all circuit parameters with their negated values. \\
The maximum absolute deviation in the resulting four electrostatic energies from the expected exact \gls{vqe} results is very low, 0.293 kcal mol$^{-1}$ for intermediate A and 0.341 kcal mol$^{-1}$ for intermediate B. This demonstrates the concurrence of our results. Nevertheless, it is noteworthy that no discernible improvement is observed with the implementation of error mitigation protocols.

\begin{figure}
    \centering
    \includegraphics[width=0.5\textwidth]{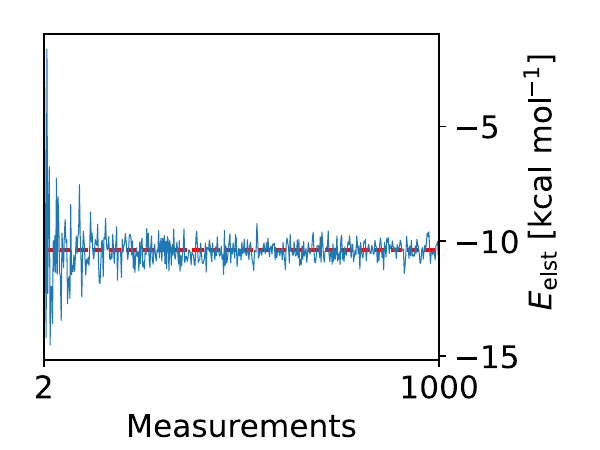}
    \caption{The convergence of the electrostatics energy for intermediate A in dependence of the number of measurements, as in Fig.~\ref{fig:experimental_results}(c).}
    \label{fig:convergence_zoom}
\end{figure}

Finally, in Fig.~\ref{fig:convergence_zoom}, we report the convergence of the electrostatics when increasing the number of measurements to construct the required \gls{1pdm} using from $2$ to $1000$ measurements. 
This is similar to Fig.~\ref{fig:experimental_results}(c) of the main text but with a focus on the first thousand measurements.
As clearly visible, at low number of measurements, the spread of values is much higher. We note that the lowest/highest possible value of electrostatics energy within the space of computational states with correct particle sector for intermediate A is $-26.98$ kcal mol$^{-1}$ and $6.80$ kcal mol$^{-1}$ respectively.
This indicates that the spectrum of $E_{\text{elst}}$ in the computational basis states is wide and that, although the average of all of them (cf. Random) takes us close to the right value, in general, a state with the wrong structure would lead to an inaccurate $E_{\text{elst}}$.

\begin{table}[h!]
\begin{tabular}{c|cc|cc}
                         & \multicolumn{2}{c|}{Intermediate A}                            & \multicolumn{2}{c}{Intermediate B}                            \\ \hline \hline
\multicolumn{1}{c|}{Random} & \multicolumn{2}{c|}{-10.096} & \multicolumn{2}{c}{-41.847 }   \\ \hline         \multicolumn{1}{c|}{VQE sim.} & \multicolumn{2}{c|}{-10.214} & \multicolumn{2}{c}{-41.528 }   \\ \hline \hline                         
\multicolumn{1}{c|}{Exp. run} & \multicolumn{1}{c|}{Plain}   & \multicolumn{1}{c|}{Mitigated} & \multicolumn{1}{c|}{Plain}   & \multicolumn{1}{c}{Mitigated} \\ \hline
\multicolumn{1}{c|}{1}   & \multicolumn{1}{c|}{-10.381(45)} & -10.409(42)                      & \multicolumn{1}{c|}{-41.699(30)} & -41.392(29)                  \\ \hline
\multicolumn{1}{c|}{2}   & \multicolumn{1}{c|}{-10.507(42)} & -10.437(43)                      & \multicolumn{1}{c|}{-41.733(30)} & -41.466(30)                     \\ \hline
\end{tabular}
\caption{This table provides an overview over the experiments ran for this study. The labeling plain corresponds to the basic setup described in the main text of this work. The experiments with the label Relabeling exploited a change of the qubit-ion assignement after each 100 shots. Moreover in these experiments, the labeling of $1$ and $0$  was exchanged. All energies are given in kcal mol$^{-1}$.}
\label{tab:additional_experiments}
\end{table}

\begin{table}[h!]
\begin{tabular}{c|cc|cc}
                         & \multicolumn{2}{c|}{Intermediate A}                          & \multicolumn{2}{c}{Intermediate B}                          \\ \hline\hline 
Random                   & \multicolumn{2}{c|}{0.191}                                   & \multicolumn{2}{c}{0.198}
\\ \hline\hline 
\multicolumn{1}{c|}{Exp. Run} & \multicolumn{1}{c|}{Plain} & \multicolumn{1}{c|}{Mitigated} & \multicolumn{1}{c|}{Plain} & \multicolumn{1}{c}{Mitigated} \\ \hline
\multicolumn{1}{c|}{1}   & \multicolumn{1}{c|}{0.544} & 0.561                           & \multicolumn{1}{c|}{0.554} & 0.547                          \\ \hline
\multicolumn{1}{c|}{2}   & \multicolumn{1}{c|}{0.573} & 0.546                           & \multicolumn{1}{c|}{0.547} & 0.572                                  \end{tabular}
\caption{Overlap between the experimental runs and the noise-less simulation of the quantum circuit. The overlap was calculated via the Bhattacharyya distance, $BC(\mathbf{p},\mathbf{q})=\sum_i \sqrt{p_iq_i}$, where $p_i$ and $p_i$ represent the frequencies of the bitstring $i$ from the experimental results and the simulation respectively.}
\label{tab:overlap_experiements}
\end{table}

\section{Calculation of the diagonal one-body Hamiltonian energy contribution}
\label{app:one_body}

To gauge the quality of our results in comparison to the supermolecular approach, we would require ground state energy calculations on the monomer and dimer systems. To estimate the energies, we would require access to the full 1- and 2-PDMs. However, we only have measured the diagonal parts of the rotated 1-PDM, see Eq.~(\ref{eq:diag_meas}).
As a proxy, we compute the expectation value of the diagonal parts of the one-body  Hamiltonian when rotated into the electrostatic potential natural orbital basis. Applying the rotations $U_{tv}$ yields

\begin{align}
    \bar{h}^A_{vv'} = \sum_{tt'} U_{vt} h_{tt'}^A U_{t'v'}
\end{align}
We then use the the measured 1-PDM from Eq.~(\ref{eq:diag_meas}) to calculate the expectation value of the diagonal terms
\begin{align}
    E_{\mathrm{diag}(\bar{h})} = \sum_v \bar{h}^A_i \bar{\gamma}^A_{vv}\,.
\end{align}
As reported in the main text, we find an error of $1.16$ kcal mol$^{-1}$, which is significantly larger than the error found for the electrostatics. 
\section{Error mitigation using Zero-Noise-Extrapolation}
\label{app:zne-extrapolation}
In order to improve the quality of the results obtained on NISQ machines, several error mitigation techniques have been developed. Typically, such methods utilize statistical procedures, that require a large number of implementations of quantum circuits, to improve the results. 

Because of this additional cost, that directly implies larger runtime on the quantum hardware, we first estimate the effect of error mitigation using a noisy hardware simulator. The dominating error source in the \textit{aqt\_marmot} hardware is depolarizing noise. Therefore, we emulate the noise using a depolarizing noise channel with a fixed error rate per gate. As mentioned in the main text, we assume error rates of $1.5\%$ and $0.3\%$ for the two qubit MS gates and the single qubit local gates. 

A suitable error mitigation method for this kind of errors is Zero-Noise-Extrapolation (ZNE). In ZNE we scale the noise by artificially increasing the circuit length. The length of the quantum circuit is thereby increased according to the noise scaling parameter $\lambda$. For $\lambda=1$ we implement the original circuit that is not altered. In general, we can use an arbitrary set of scaling parameters. After the implementation of the circuits with variable lengths, e.g. $\lambda=[1,2,3]$, we extrapolate the behavior of the device for the zero noise case $\lambda=0$. For the extrapolation we define a figure of merit, typically an expectation value, and a fitting function. Depending on the scaling of the expectation value with noise we choose a suitable fitting function. This can be a linear, a polynomial, an exponential or any other function that is able to resemble the noise scaling. 

We implement ZNE using the open source python toolkit \textit{mitiq}~\cite{mitiq}. As the noise scaling function, we choose to use the unitary folding strategy \verb|fold_gates_at_random|, that is described in detail in~\cite{mitiq}. The scaled circuits resemble the same unitary as the original circuit, where the total number of gates $n_\lambda\approx n_0 \lambda$ is roughly the number of gates of the original circuit $n_0$ multiplied by the scaling factor $\lambda$. 

As a testbed, we use the 8-qubit circuit to estimate the electrostatic energy of the quantum circuit corresponding to intermediate A, that contains a gate count of $(R_\mathrm{Z}, 168), (R_\mathrm{X}, 111), (R_\mathrm{XX}, 63)$ in the original version. ZNE is performed using the noisy simulator with scaling parameters $\lambda=[1,2,3]$, where each of the three circuits is implemented a total of 60000 times (300 noisy simulations with 200 shots each). Since the $R_\mathrm{XX}$ gate is not supported in \textit{mitiq}, we transpile the circuit into the gate set $R_\mathrm{Z}, R_\mathrm{X}, CX$ (a controlled-X or controlled-NOT gate). To prevent the removal of the additional gates in the scaled circuits, we use the \verb|optimization_level=1| option in the Qiskit transpiler. Therefore the gate count is slightly different compared to the original circuit $(R_\mathrm{Z}, 139), (R_\mathrm{X}, 121), (R_\mathrm{XX}, 63)$ for $\lambda=1$, $(R_\mathrm{Z}, 140), (R_\mathrm{X}, 239), (R_\mathrm{XX}, 122)$ for $\lambda=2$ and $(R_\mathrm{Z}, 140), (R_\mathrm{X}, 372), (R_\mathrm{XX}, 189)$ for $\lambda=3$.
\begin{figure}[t]
    \centering
    \includegraphics[width=0.45\linewidth]{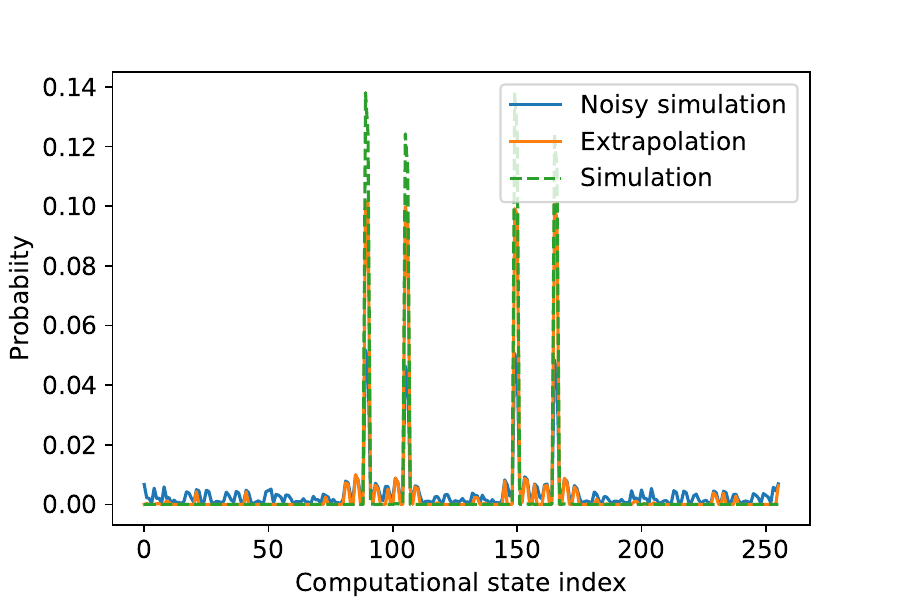}
        \includegraphics[width=0.45\linewidth]{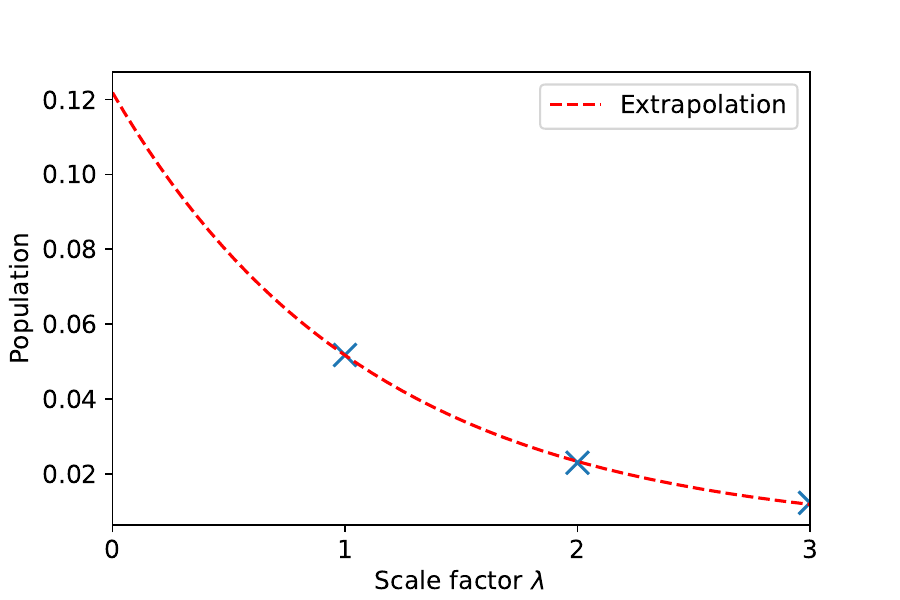}

    \caption{(left) A comparison of the output of the noise-less and the noisy output statistics together with results obtained from extrapolating the frequencies to zero noise. (right) The output of the extrapolation of the computational state with highest frequency.}
    \label{fig:zne}
\end{figure}

To extrapolate the electrostatic interaction energy, which is calculated through the 1-PDM given in Eq.~(\ref{eq:diag_meas}), we extrapolate the frequencies of the computational states, using the following fit function
\begin{align}
    f(\lambda)=a+b\exp{(-c\lambda)}
\end{align}
with an additional data point of $1/2^8$ for $\lambda=10^5$, see Fig.~\ref{fig:zne}(right) for the extrapolation of the computational state with highest frequency. All computational states with an extrapolated frequency $<0$ are set to $0$. The error of the extrapolated frequency is obtained by $\sigma^{(ZNE)}_{\mathrm{freq.}}=\sqrt{\sigma_a^2+\sigma_b^2}$, where $\sigma_a$ and $\sigma_b$ are the standard deviations of $a$ and $b$ obtained from the fit. In Fig.~\ref{fig:zne}(left), we show a comparison between the noise-less, the noisy and the normalized extrapolated results. As visible, the extrapolation boosts the overlap w.r.t. the noise-less simulation. In fact, the Bhattacharyya distance of the noise-less simulation and the noisy simulation increases from $0.63$ to $0.90$ after extrapolation. We then utilize these extrapolated frequencies to reconstruct the 1-PDM, as per Eq.~(\ref{eq:diag_meas}), and subsequently compute the electrostatics. We propagate the errors from the extrapolation into the 1-PDM and into the electrostatics using standard Gaussian error propagation.
Our findings indicate that the error in the electrostatics relative to the noise-free simulation did improve as anticipated from $\Delta{E}_\mathrm{elst.}^{\mathrm{noisy}}= 0.144$ kcal mol$^{-1}$ to $\Delta{E}_\mathrm{elst.}^{\mathrm{ext.}}= 0.076\pm0.126$ kcal mol$^{-1}$. However, we note that the  of the extrapolation is within the original value.   
\end{document}